\documentclass[12pt]{article} 
\usepackage{graphicx}
\usepackage{amsmath}
\usepackage{amssymb}
\usepackage{cite}

\makeatletter
 
  \@addtoreset{equation}{section}
 \makeatother

\newcommand{\bea}{\begin{eqnarray}}
\newcommand{\be}{\begin{equation}}
\newcommand{\ee}{\end{equation}}
\newcommand{\eea}{\end{eqnarray}}
\newcommand{\beann}{\begin{eqnarray*}}
\newcommand{\eeann}{\end{eqnarray*}}
\newcommand{\nn}{\nonumber}
\newcommand{\ba}{\begin{array}}
\newcommand{\ea}{\end{array}}

\DeclareMathOperator{\Tr}{Tr}

\newcommand{\del}{\partial}

\newcommand{\cO}{{\mathcal O}}
\newcommand{\cD}{{\mathcal D}}
\newcommand{\cN}{{\mathcal N}}

\newcommand{\hQ}{\hat{Q}}

\newcommand{\link}[1]{{\left\langle #1 \right\rangle}}

\title{
Topologically Twisted  $N=(2,2)$ Supersymmetric Yang-Mills Theory
on Arbitrary Discretized Riemann Surface
}

\author{ %{}\\
So Matsuura${}^1$, Tatsuhiro Misumi${}^1$ and Kazutoshi Ohta${}^2$\\[0.7cm]
{\em ${}^1$ Department of Physics in Hiyoshi Campus, Keio University, } \\
{\em 4-1-1 Hiyoshi, Yokohama, 223-8521, Japan} \\
{\em ${}^2$ Institute of Physics, Meiji Gakuin University, Yokohama 244-8539, Japan}}
\date{\today}							% Activate to display a given date or no date

\begin{document}
\maketitle

\vspace*{1.5cm}

\begin{center}
{\bf Abstract}
\end{center}

We define supersymmetric Yang-Mills theory 
on an arbitrary two-dimensional lattice (polygon decomposition)
while preserving one supercharge.
When a smooth Riemann surface $\Sigma_g$ with genus $g$
emerges as an appropriate continuum limit 
of the generic lattice, 
the discretized theory 
becomes a topologically twisted 
$\cN=(2,2)$ supersymmetric Yang-Mills theory on $\Sigma_g$. 
If we adopt the usual square lattice as a special case of the discretization, 
our formulation is identical with Sugino's lattice model.
Although the tuning of parameters is generally
required while taking the continuum limit, the number of necessary parameters is at most
two because of the gauge symmetry and the supersymmetry.
In particular, we do not need any fine-tuning
if we arrange the theory so as to possess an extra global $U(1)$ symmetry
 ($U(1)_{R}$ symmetry) which rotates the scalar fields.

%We do not need any fine-tuning in taking the continuum limit 
%when the theory has the $U(1)$ symmetry, 
%while we need one-parameter (two-parameter) tuning 
%for the gauge group 
%$SU(N)$ ($U(N)$) when the $U(1)$ symmetry is absent. 
\newpage

%%%%%%%%%%%%%%%%%%%%%%%%%%%%%%%%%%%%%%%%%%%%
\section{Introduction}
Since the middle of  the 1980s, after the first success of numerical QCD simulations based on lattice regularization, 
extension of the lattice technique to supersymmetric gauge theories has been pursued with great interest \cite{Elitzur:1982vh, Banks:1982ut,Ichinose:1982ug,Bartels:1983wm}.
The hindrance encountered there was the fact that the regularization 
breaks the Poincar\'e invariance to its discrete subgroup 
and the supersymmetry cannot be straightforwardly realized on the lattice.
To date, however, several lattice formulations of supersymmetric gauge theories 
have been developed by bypassing this difficulty. 
In particular, for one or two-dimensional theories with extended supersymmetries, 
there are such lattice formulations that are free from fine-tuning 
in taking the continuum limit thanks to partially preserved supercharges on the lattice. 

In \cite{Kaplan:2002wv,Cohen:2003xe,Cohen:2003qw,Kaplan:2005ta,Endres:2006ic,Giedt:2006dd,Matsuura:2008cfa,Catterall:2003wd,DAdda:2004jb,Nagata:2008zz,Joseph:2013jya, Unsal:2006qp,Catterall:2007kn,Damgaard:2007eh}, 
some of  the supercharges are exactly preserved on a hypercubic lattice 
by applying the so-called orbifolding 
procedure to supersymmetric matrix theory (mother theory)\footnote{
For a review, see \cite{Catterall:2009it}.}.
In these formulations, the bosonic link variables are not unitary but complex matrices, which restricts gauge groups  to $U(N)$ rather than $SU(N)$. In numerical simulations, therefore, we must introduce a large mass in the $U(1)$ part of the complex link variables in order to fix the lattice spacing and take care of the fermionic zero modes in computing the Dirac matrix~\cite{Kanamori:2008bk,Hanada:2009hq,Catterall:2010fx}. 
In \cite{Sugino:2003yb,Sugino:2004qd,Sugino:2004uv,Sugino:2006uf,Sugino:2008yp,Kikukawa:2008xw}, 
the authors discretized topologically twisted gauge theories while preserving one or two supercharges. 
In these formulations, 
lattice gauge fields are expressed by compact link variables on the hypercubic lattice, 
as in conventional lattice gauge theories and we can choose the gauge group $SU(N)$,
which will be more convenient for numerical 
simulations~\cite{Suzuki:2007jt,Kanamori:2007ye,Kanamori:2007yx}. 
In addition, the problem of the vacuum degeneracy of lattice gauge fields 
pointed out in these models \cite{Sugino:2004qd} 
has recently been solved without using an admissibility condition \cite{Matsuura:2014pua}.

As for three- and four-dimensional supersymmetric theories, 
apart from the formulations \cite{Maru:1997kh,Giedt:2009yd} 
with exact chiral symmetry enabling the whole supersymmetry restoration in the continuum limit, 
lattice-regularized gauge theories require parameter tunings in taking the continuum limit 
even if part of supersymmetry is exactly preserved, 
since the symmetries on the lattice are generally insufficient 
to forbid relevant operators that break the rest symmetries\footnote{
As another approach to circumvent this issue, 
four-dimensional $\cN=4$ supersymmetric Yang-Mills theory in the planar limit 
can be obtained by using a large-$N$ reduction technique 
which has been extensively studied from both the theoretical and numerical points of view
\cite{Ishii:2008ib,Ishiki:2008te,Ishiki:2009sg,Ishiki:2011ct,Honda:2013nfa}. 
As for theories with finite rank gauge group, 
a hybrid regularization has been proposed for four-dimensional $\cN=2, \,4$ supersymmetric Yang-Mills 
theories~\cite{Hanada:2010kt,Hanada:2010gs,Hanada:2011qx}, where 
two different discretizations by lattice and matrix~\cite{Berenstein:2002jq,Das:2003yq,Myers:1999ps}
are combined. 
For another numerical approach to $\cN=4$ SYM, 
see \cite{Catterall:N=41,Catterall:N=42,Catterall:N=43,Catterall:N=44,Catterall:N=45}. 
}.

As a common feature of lattice gauge theories so far, 
no attention has been paid to the topology of the spacetime. 
Indeed, all the previous lattice formulations of supersymmetric theories are discretized 
on a periodic hypercubic lattice;  thus, the topology is always torus. 
Although this is natural because the main interest in conventional lattice gauge theories is 
in the UV nature, where the topology of the spacetime is usually irrelevant,
it is also true that the topology is sometimes quite important for supersymmetric gauge 
theories especially in the context of topological field theory \cite{Witten:1988ze}. 
The importance of such theories has recently been increasing again, 
in relation to the height of the localization technique in supersymmetric gauge theories 
\cite{Pestun:2007rz}.

In this paper, we consider topologically twisted two-dimensional $\cN=(2,2)$ 
supersymmetric Yang-Mills theory on a generic Riemann surface. 
We discretize the Riemann surface to an arbitrary lattice (polygons) and propose a way to 
define the supersymmetric gauge theory on it while preserving a supercharge. 
We show that we can define the theory on any decomposition of the two-dimensional surface 
and 
the tree-level continuum limit reproduces the continuum theory. 
We see that, 
if we consider the usual square lattice as a special case of discretization, 
our formulation coincides with Sugino's formulation 
\cite{Sugino:2003yb,Sugino:2004qd,Sugino:2004uv,Sugino:2006uf}. 
We discuss that there are two types of theories depending on the hermiticity 
of the scalar fields: theories with and without an extra global $U(1)$ symmetry
 ($U(1)_{R}$ symmetry). 
If the theory has this symmetry, we can take the continuum limit without any fine-tuning, 
while we need one-parameter (two-parameter) tuning in taking the continuum limit 
if the theory does not have this symmetry and the gauge group is $SU(N)$ ($U(N)$).

This paper is organized as follows. 
In the next section, we briefly review the continuum topologically twisted two-dimensional 
$\cN=(2,2)$ supersymmetric Yang-Mills theory on a curved background. 
In section 3, we define the theory on a general lattice and discuss 
the continuum limit and possible radiative corrections. 
The section 4 is devoted to the conclusion and discussion. 
In appendix \ref{proof}, we calculate the continuum limit of a face variable in detail. 

%%%%%%%%%%%%%%%%%%%%%%%%%%%%%%%%%%%%%%%%%%%%
\section{Continuum two-dimensional ${\cal N}=(2,2)$ supersymmetric Yang-Mills theory}

%\subsection{continuum theory on a flat spacetime}
We start with the two-dimensional ${\cal N}=(2,2)$ supersymmetric Yang-Mills theory 
on a flat Euclidean spacetime, which is obtained from a dimensional reduction of 
four-dimensional ${\cal N}=1$ supersymmetric Yang-Mills theory: 
\begin{align}
	S=\frac{1}{2g_{2d}^2}\int d^2x \, \Tr \biggl\{
 &\frac{1}{2}F_{\mu\nu}^2+\left(\cD_\mu \Phi\right)\left(\cD_\mu \bar\Phi\right)
 +\frac{1}{4}\left[\Phi,\bar\Phi\right]^2 \nn \\
 &+i \bar\Psi \Gamma_\mu \cD_\mu \Psi -\frac{1}{2} \bar\Psi \Gamma_+ \left[\bar\Phi,\Psi\right]
 -\frac{1}{2}\bar\Psi\Gamma_- \left[ \Phi,\Psi \right]
 \biggr\},
 \label{action1}
\end{align}
where $\mu,\nu=1,2$, 
$\Gamma_\mu$ and $\Gamma_{\pm}=\Gamma_3\pm i\Gamma_4$ are four-dimensional 
Dirac matrices satisfying $\{ \Gamma_M, \Gamma_N \}=-2\delta_{MN}$ $(M,N=1,\cdots,4)$, 
$\Psi$ is a four-component spinor, 
$\bar\Psi=-i\Psi^T \Gamma_4$, 
$F_{\mu\nu}$ is the field strength of a gauge field $A_\mu$, 
and $\Phi$ and $\bar\Phi$ are scalar fields. 
We assume that the gauge group $G$ is $U(N)$ or $SU(N)$ in the following. 
%Usually $\bar\Phi$ is considered as the hermitian conjugate of $\Phi$, 
%but in the following we regard that $\Phi$ and $\bar\Phi$ are independent 
%hermitian matrices. 

We fix the notation of the gamma matrices by 
\begin{align}
 \Gamma_1=\left(
 \begin{matrix}
 i\sigma_3 & \\
 & i\sigma_3
 \end{matrix}\right), \quad
\Gamma_2=\left(
 \begin{matrix}
 i\sigma_1 & \\
 & i\sigma_1
 \end{matrix}\right), \quad
 \Gamma_3=\left(
 \begin{matrix}
 & -\sigma_2  \\
  \sigma_2
 \end{matrix}\right), \quad
 \Gamma_4=\left(
 \begin{matrix}
 & -i\sigma_2  \\
  -i\sigma_2
 \end{matrix}\right), 
\end{align}
and express the components of the spinor $\Psi$ as
\begin{equation}
 \Psi=\left( \lambda_1, \lambda_2, \chi, \eta/2 \right)^T. 
\end{equation}
Then (\ref{action1}) reduces to
\begin{align}
	S=\frac{1}{2g_{2d}^2}\int d^2x \,  \Tr \biggl\{
 &\frac{1}{2}F_{\mu\nu}^2+\left(\cD_\mu \Phi\right)\left(\cD_\mu \bar\Phi\right)
 +\frac{1}{4}\left[\Phi,\bar\Phi\right]^2 \nn \\
 &+i\eta\cD_\mu\lambda_\mu + 2i \chi\left(\cD_1\lambda_2-\cD_2\lambda_1\right)
 +\lambda_\mu\left[\bar\Phi, \lambda_\mu \right]
 -\chi\left[\Phi,\chi\right]
 -\frac{1}{4}\eta \left[ \Phi, \eta \right]
 \biggr\}.
 \label{action2}
\end{align}
We see that (\ref{action1}) (and of course (\ref{action2})) is invariant under 
the supersymmetric transformation,
\begin{equation}
\begin{array}{l}
 \delta \Phi = -i\bar\xi \Gamma_+ \Psi, \quad
 \delta \bar\Phi = -i\bar\xi \Gamma_- \Psi, \quad
 \delta A_\mu=-i\bar\xi \Gamma_\mu \Psi, \\
 \delta \Psi = -F_{12} \Gamma_{12} \xi -\frac{1}{2}\left( \cD_\mu \bar\Phi \right)
 \gamma_{\mu+} \xi -\frac{1}{2}\left( \cD_\mu \Phi \right) \Gamma_{\mu -} \xi
 -\frac{i}{4}\left[\Phi,\bar\Phi\right] \Gamma_{+-} \xi,
\end{array}
\end{equation}
where $\xi$ is a four-component spinor parameter and 
$\Gamma_{MN}\equiv\frac{1}{2}\left[\Gamma_M, \Gamma_N\right]$.  

Now let us consider a specific SUSY transformation associated with the parameter 
$\xi=\left(0,0,0,\epsilon\right)^T$ and define the corresponding supercharge $\hQ$ as\footnote{
Here we have put a hat on the supercharge $Q$ 
in order to distinguish it from the one
appeared in the discretized theory in the next section.}
\begin{equation}
 \delta \phi \equiv -i \epsilon \left( \hQ \phi \right), 
\end{equation}
for an arbitrary field $\phi$. 
We can read off the $\hQ$-transformation of the fields as 
\begin{equation}
\begin{array}{lcl}
 \hQ\Phi = 0, && \\
 \hQ \bar\Phi = \eta, && \hQ\eta = [\Phi, \bar\Phi],\\
 \hQ A_\mu = \lambda_\mu, &&  \hQ \lambda_\mu = i D_\mu \Phi,\\
\hQ Y = [\Phi, \chi], && \hQ \chi = Y,
\end{array}
\label{cont SUSY}
\end{equation}
where $Y$ is an auxiliary field. Then the action (\ref{action2}) can be expressed 
in the $\hQ$-exact or 
topologically twisted form \cite{Witten:1988ze,Witten:1990bs} by 
\begin{align}
	S=\hQ\frac{1}{2g_{2d}^2}\int d^2x \Tr \left[ \frac{1}{4}\eta \left[ \Phi, \bar\Phi \right] 
-i \lambda^\mu D_\mu \bar\Phi +\chi\left(Y - 2i F_{12} \right) \right]. 
\label{flat action}
\end{align}
  It is important that the $\hQ^2$ is equal to the infinitesimal gauge transformation with
a parameter $\Phi$. 
Since $\hQ$ is acting on a gauge-invariant expression in the action (\ref{flat action}), 
the $\hQ$-invariance of the action is manifest. 

%\subsection{topological theory on a curved background}

We next extend the above theory to that on a curved background. 
One of the motivations for considering topological twist is 
to preserve a partial supersymmetry in a curved background \cite{Bershadsky:1995qy}.
The supersymmetry we usually use 
is completely broken on a curved background because 
there is in general no covariantly constant spinor. 
However, by twisting the local Lorentz symmetry with R symmetry, 
there can appear ``scalar supercharges'' which are preserved in any curved background.  
The supercharge $\hQ$ in (\ref{flat action}) becomes the scalar supercharge
as it is, 
and thus we can define topological Yang-Mills theory on the curved spacetime
while keeping $\hQ$ as
\begin{align}
	S=\hQ\frac{1}{2g_{2d}^2}\int_{\Sigma_g} d^2x \sqrt{g} \Tr \left[ \frac{1}{4}\eta \left[ \Phi, \bar\Phi \right] 
-i g^{\mu\nu} \lambda_\mu D_\nu \bar\Phi +\chi\left(H - 2i f \right) \right], 
\label{curved action}
\end{align}
where 
the covariant derivative $D_\mu$ now includes not only the gauge field but also the spacetime 
connection, 
$\hQ$ is the same as in (\ref{cont SUSY}), 
$\Sigma_g$ is an oriented or unoriented two-dimensional manifold with the metric $g_{\mu\nu}$%
\footnote{
$\Sigma_g$ can have even boundaries. In that case, we take the free boundary condition for simplicity. 
}
and $f(x)=\frac{1}{2}\frac{\epsilon^{\mu\nu}}{\sqrt{g(x)}}F_{\mu\nu}(x)$ is the Poincar\'e dual of 
the field strength. 
Because of the deformation of the background, the other three supersymmetries are broken in general. 

Here we make some comments. 
{}First, the operations of twisting and curving do not commute. 
The action (\ref{curved action}) is obtained by twisting the theory on the flat spacetime 
followed by curving the background. 
This theory differs from the one obtained by first curving the background 
followed by twisting (or renaming the fermionic fields). 
In the following section, we discretize the former (topological) theory. 
Therefore, even if we take the continuum limit, we do not obtain the latter (physical) theory. 
We note that 
it does not conflict with the fact that the continuum limit of Sugino's lattice formulation 
is the physical supersymmetric gauge theory \cite{Sugino:2003yb,Sugino:2004qd,Sugino:2004uv,Sugino:2006uf}. 
This is because Sugino's formulation is defined on a flat spacetime where the physical 
theory and the topological theory coincide and twisting is merely a renaming of the fields. 

Second, 
we can choose the hermiticity of the scalar fields $\Phi(x)$ and $\bar\Phi(x)$. 
They are usually regarded as hermitian conjugate with each other from the construction; 
they are originally related to the components of the gauge fields of the four-dimensional theory 
as $\Phi=A_3+iA_4$ and $\bar\Phi=A_3-iA_4$. 
In this case, the theory possesses $U(1)_R$ symmetry, 
\begin{equation}
\begin{array}{lll}
 \Phi\to e^{2i\alpha}\Phi, & \bar\Phi\to e^{-2i\alpha}\bar\Phi, & 
A_\mu\to A_\mu,\\
 \eta\to e^{-i\alpha}\eta, &
  \lambda_\mu \to e^{i\alpha} \lambda_\mu, &
  \chi\to e^{-i\alpha}\chi.
  \end{array}
 \label{U1-R}
\end{equation}
On the other hand, as often adopted in the context of the topological field theory, 
we can instead regard $\Phi(x)$ and $\bar\Phi(x)$ as independent hermitian variables. 
As a result, it is impossible to impose the $U(1)$ rotation (\ref{U1-R}).
This choice completely changes the theory. 
{}For example, the expectation value $\langle \int d^2x \sqrt{g(x)} \Tr( \Phi(x)^n ) \rangle$ is zero 
in the former theory because of the $U(1)_R$ symmetry (\ref{U1-R}) 
but it takes some non-trivial value in the latter theory. 
We can consider both theories depending on the purpose 
and can use the same discretization, explained in the next section.

%%%%%%%%%%%%%%%%%%%%%%%%%%%%%%%%%%%%%%%%%%%%
\section{${\cal N}=(2,2)$ supersymmetric Yang-Mills theory on an arbitrary discretized Riemann surface}

In this section, we discretize the continuum theory described in the previous section on 
a given decomposition of the two-dimensional surface, i.e., a set of sites, links and faces. 
As mentioned in the previous section, we can use the same discretization 
if we regard the scalar field $\Phi$ as either complex or hermitian 
so we do not specify it in constructing the discretized formulation.  
We will see, however, that this choice is crucial in considering radiative corrections. 

%%%
\subsection{Definition of the model}
A polygon decomposition of the two-dimensional surface consists of a set of sites $S$, links $L$ and faces $F$, respectively: 
\begin{align}
 S&\equiv\{s|s=1,\cdots,N_S\},\nn \\
 \label{data} 
 L&\equiv\{\link{st} |s,t\in S\},  \\
 F&\equiv\{(s_1,\cdots,s_n)| s_1,\cdots,s_n\in S,\ (s_i,s_{i+1})\in L\ {\rm or}\ (s_{i+1},s_i)\in L\}, \quad
 (s_{n+1}\equiv s_1), \nn
\end{align}
where $N_S$ is the number of sites, 
a link $\link{st}$ possesses a direction from $s$ to $t$,
and a face $(s_1,\cdots,s_n)$ is a surface surrounded by the links
$\link{s_i \, s_{i+1}}$  $(i=1,\cdots,n)$%
\footnote{ 
Only the sites $s_i$ and $s_{i+1}$ $(i=1,\cdots,n)$ must be connected by a link. }.
We sometimes call the first site $s_1$ of the face $f\equiv (s_1,\cdots,s_n)$ as 
the representative point (site) of the face $f$. 
This is apparently a generalization of the usual square lattice which is given by the data, 
\begin{align}
S&=\{ \vec{X}=(x,y) | 1\le x \le L_x,\ 1\le y \le L_y \}, \nn \\
L&=\left\{ \link{\vec{X}\, \vec{X}+\hat{x}}, \link{\vec{X}\,\vec{X}+\hat{y}} | \vec{X}\in S\right\}, \\
F&=\left\{ \left( \vec{X}, \vec{X}+\hat{x}, \vec{X}+\hat{x}+\hat{y}, \vec{X}+\hat{y} \right) | \vec{X} \in S \right\}. \nn
\label{square lattice}
\end{align}

We next consider the following ``fields'' associated with 
the sites, links and faces 
of a given decomposition, respectively: 
\begin{align}
 \Phi_s, \bar\Phi_s, \eta_s &:\ \text{site variables}\ (s\in S), \nn \\
 U_{st}, \Lambda_{st} &:\ \text{link variables}\ (\link{st}\in L),  \\
 Y_f, \chi_f &:\ \text{face variables}\ (f\in F), \nn
\end{align}
where $\Phi_s$, $\bar\Phi_s$, $U_{st}$ and $Y_f$ are bosonic variables 
and $\eta_s$, $\Lambda_{st}$ and $\chi_f$ are fermionic variables. 
We assume that the site variables $\Phi_s$, $\bar\Phi_s$ and $\eta_s$ live on the site $s$, 
the link variables $U_{st}$ and $\Lambda_{st}$ live on the link $\link{st}$, 
and the face variables $Y_{f}$ and $\chi_f$ live on the representative point of the face $f$.
%\footnote{Formally, we can define a ``field on a face'' like 
%$Y_{(stu)}\equiv Y_{s}U_{st} U_{tu} U_{us}$ for a triangle face $(stu)$. 
%(The generalization to an arbitrary face is trivial. )
%Although we do not use it in this paper, we can also change the representative point like 
%$Y_{(stu)}\equiv U_{st} Y_t U_{tu}U_{us}$. 
%Note that the exchange $Y_s$ and $Y_{stu}$ corresponds to taking the Hodge dual. 
%}
We often express the link fermion $\Lambda_{st}$ as 
\begin{equation}
 \Lambda_{st} \equiv \lambda_{st} U_{st}, 
\end{equation}
where $\lambda_{st}$ lives on the site $s$. 
We assume that $U_{st}\in G$ and the other fields including $\lambda_{st}$ 
are in the adjoint representation of $G$. 
{}For a given link $\link{st}$, we sometimes use the notation $U_{ts}\equiv U_{st}^{-1}$. 
Then the gauge transformations of the fields are
\begin{align}
 \Phi_s &\to g_s \Phi_s g_s^{-1}, &
 \bar\Phi_s &\to g_s \bar\Phi_s g_s^{-1}, &
 \eta_s &\to g_s \eta_s g_s^{-1},  \nn\\
 U_{st} &\to g_s U_{st} g_t^{-1}, &
 \Lambda_{st} &\to g_s \Lambda_{st} g_t^{-1},  \\
 Y_f &\to g_f Y_f g_f^{-1}, &
 \chi_f &\to g_f \chi_f g_f^{-1}, \nn
\end{align}
where $g_s \in G$ $(s\in S)$ 
and we have used the same symbol $f$ to describe a face and the representative point 
in the last line. 
It is easy to see that $\lambda_{st}$ transforms as 
$\lambda_{st} \to g_s \lambda_{st} g_s^{-1}$ under the gauge transformation. 

Corresponding to the SUSY transformation (\ref{cont SUSY}), we consider the following 
transformation of the fields on the general lattice: 
\begin{equation}
\begin{array}{lcl}
 Q\Phi_s = 0, && \\
  Q\bar\Phi_s = \eta_s, && Q\eta_s=[\Phi_s, \bar\Phi_s], \\
 Q U_{st}  = i \lambda_{st} U_{st}, &&  Q \lambda_{st} = i\left( U_{st}\Phi_t U_{st}^{-1} - \Phi_s + \lambda_{st}\lambda_{st} \right),\\
Q Y_f = [\Phi_f, \chi_f], && Q\chi_f = Y_f.
 \label{SUSY}
\end{array}
\end{equation}
Note that the third line can be rewritten as 
\begin{equation}
 Q U_{st}=i \Lambda_{st}, \quad Q \Lambda_{st} = i\left( U_{st} \Phi_t - \Phi_s U_{st} \right),
\end{equation}
in terms of $\Lambda_{st}$ instead of $\lambda_{st}$.
It is easy to see that $Q^2$ is equal to the infinitesimal gauge transformation 
with the parameter $\Phi_s$; thus, $Q$ is nilpotent if it acts on a gauge-invariant 
expression. 
Using this supercharge, we define the action, 
\begin{align}
 S&= S_S+S_L+S_F \nn \\ 
  &\equiv Q \sum_{s\in S} \alpha_s \Xi_s 
  + Q \sum_{\link{st}\in L} \alpha_{\link{st}} \Xi_{\link{st}} 
  + Q \sum_{f\in F} \alpha_f \Xi_f, 
 \label{lat action}
\end{align}
with
\begin{align}
\label{pre action site}
 \Xi_s &\equiv \frac{1}{2g_0^2}\Tr \biggl\{ \frac{1}{4} \eta_s [\Phi_s,\bar\Phi_s] \biggr\}, \\ 
 \label{pre action link}
 \Xi_{\link{st}} &\equiv \frac{1}{2g_0^2}\Tr \biggl\{ -i \lambda_{st} \left(
 U_{st} \bar\Phi_t U_{st}^{-1} - \bar\Phi_s \right) \biggr\}, \\
 \Xi_f &\equiv \frac{1}{2g_0^2}\Tr \biggl\{ \chi_f \left( Y_f - i\beta_f \mu(U_f) \right) \biggr\},
 \label{pre action face}
\end{align}
where $\alpha_s$, $\alpha_{\link{st}}$, $\alpha_f$ and $\beta_f$ 
are constants that will be fixed later so that the theory 
has an appropriate continuum limit, 
$\mu(U_f)$ is given by \cite{Matsuura:2014pua}
\begin{equation}
 \mu(U_f) = 
 \begin{cases}
 2i \Bigl[\left( U_f-U_f^{-1} \right)^{-1}\left(2-U_f-U_f^{-1}\right)  \\
\hspace{2cm} + \left(2-U_f-U_f^{-1}\right)  \left( U_f-U_f^{-1} \right)^{-1} \Bigr] 
\quad &{\rm for}\ G=U(N),  \\
\displaystyle \frac{2i}{M}
\Bigl[
 \left( U_f^M - U_f^{-M} \right) \left( 2 - U_f^M - U_f^{-M} \right) \\
\hspace{2cm}
+ \left( 2 - U_f^M - U_f^{-M} \right) \left( U_f^M - U_f^{-M} \right) 
 \Bigr] \quad & {\rm for}\ G=SU(N), 
\end{cases}
\end{equation} 
with $2M>N$, 
and $U_f$ is the ``plaquette variable'' defined by  
\begin{equation}
 U_f\equiv \prod_{i=1}^n U_{s_i s_{i+1}}, 
\end{equation}
for $f=(s_1,\cdots,s_n)$. 
Note that the form of $\mu(U_f)$ is determined in order that the theory 
possesses unique vacuum at $U_f=1$ (see \cite{Matsuura:2014pua} for details). 
The explicit expression of the action is
\begin{equation}
 S=S_b+S_f, 
 \label{action}
\end{equation}
with 
\begin{align}
\label{bosonic action}
 S_b= &\frac{1}{2g_0^2}\sum_{s\in S}\alpha_s \Tr\left\{ \frac{1}{4}[\Phi_s,\bar\Phi_s]^2 \right\} \nn \\
 &+ \frac{1}{2g_0^2}\sum_{\link{st}\in L}\alpha_{\link{st}} \Tr\biggl\{
 (U_{st}\Phi_t U_{st}^{-1} - \Phi_s) (U_{st}\bar\Phi_t U_{st} - \bar\Phi_s) \biggr\} \nn \\
&+ \frac{1}{2g_0^2}\sum_{f\in F}\alpha_f \Tr \biggl\{ Y_f ( Y_f - i \beta_f \mu(U_f)) \biggr\}, \\
 S_f= &\frac{1}{2g_0^2}\sum_{s\in S}\alpha_s \Tr\left\{ -\frac{1}{4} \eta_s [\Phi_s, \eta_s] \right\} \nn \\
 &+\frac{1}{2g_0^2}\sum_{\link{st}\in L }\alpha_{\link{st}} \Tr\biggl\{ -i \lambda_{st}(U_{st}\eta_t U_{st}^{-1} - \eta_s )
 - \lambda_{st}\lambda_{st}( U_{st} \bar\Phi_t U_{st}^{-1} + \bar\Phi_s ) \biggr\} \nn \\
 &+\frac{1}{2g_0^2}\sum_{f\in F}\alpha_{f} \biggl\{ -\chi_f [\Phi_f,\chi_f] + i\beta_f \chi_f \Bigl(Q\mu(U_f)\Bigr) \biggr\}.
\end{align}
If we consider the torus discretization corresponding to the square lattice (\ref{square lattice})
and set $\alpha_s=\alpha_\link{st}=\alpha_f=\beta_f=1$, 
this action reproduces that of the lattice formulation of two-dimensional $\cN=(2,2)$ 
supersymmetric Yang-Mills theory given in \cite{Sugino:2003yb,Matsuura:2014pua}.

We make a comment before closing this subsection. 
The construction of the discretized theory given above is based 
on abstract data (\ref{data}) which includes such polygons 
that cannot be interpreted as a discretization of any Riemann surface%
\footnote{The 3D cubic lattice is a typical example.}. 
Since our main purpose in this paper is to discretize the two-dimensional 
topological gauge field theory, 
we will implicitly restrict the polygons to discretized Riemann surfaces in the next section. 
However, it is worth noting that our construction is applicable 
to a wider class of discretized objects in principle. 

%%%
\subsection{Classical continuum limit}
% tree-level continuum limit
Let us next consider the tree-level continuum limit. 
To this end, 
we assume that the given decomposition is sufficiently fine %(large)
to approximate a Riemann surface $\Sigma_g$. 
We first define the ``lattice spacing'' through the relation, 
\begin{equation}
 a^2 N_F = \int_{\Sigma_g} d^2x \sqrt{g(x)}, 
 \label{lat space}
\end{equation}
where $N_F$ is the number of faces. 
In other words, $a^2$ is equal to the average area of the faces. 
The continuum limit is defined by the limit of $a\to 0$ and $N_F\to\infty$ while fixing
(\ref{lat space}). 
We also define the area of each face as
\begin{equation}
 a^2 A_f = \int_{\sigma_f} d^2 x \sqrt{g(x)},  
\end{equation}
where the integration is taken over the region (simplex) $\sigma_f$ corresponding to the face $f$. 
In particular, we see 
\begin{equation}
 a^2 \sum_{f\in F} A_f \to  \int_{\Sigma_g} d^2x \sqrt{g(x)}, 
\end{equation}
in the continuum limit.

Since we assume that the given decomposition sufficiently well approximates 
the Riemann surface $\Sigma_g$, 
we can identify the index $s$ of a site with a two-dimensional coordinate $x_s$. 
Then, corresponding to the link $\link{st}$, we can define a covariant vector, 
\begin{equation}
 e^\mu_{st} \equiv \frac{1}{a} \left( x_t^\mu - x_s^\mu \right), 
\end{equation}
where $x_s$ and $x_t$ are 
the two-dimensional coordinates corresponding to the sites $s$ and $t$, respectively. 
Here let $L_f$ denote a set of links that construct the face $f$. 
{}From the definition of the continuum limit, it is natural to identify a face 
as a tangent space of the Riemann surface. 
Thus we assume that all the vectors $e^\mu_{st}$ for $\link{st}\in L_f$ are 
in the same two-dimensional plane.

Here we should note that all the fields
on a general lattice are defined as 
dimensionless quantities, 
thus we must supply appropriate powers of $a$ in order to define
the corresponding continuum fields. 
We must also require that the correspondence must be consistent with the $Q$-transformation. 
{}From these requirements, it is natural to consider the following correspondence
between the discrete and continuum fields:
\begin{equation}
\begin{array}{l}
 \Phi_s=a \Phi(x_s), \quad \bar\Phi_s= a \bar\Phi(x_s), \quad 
 \eta_s=a^{\frac{3}{2}}\eta(x_s),  \\
  U_{st}=e^{iae^\mu_{st} A_\mu(x_s+\frac{a}{2}e_{st}^\mu)},  \\
 \lambda_{st}=a^{\frac{3}{2}}e^{\frac{i}{2}ae^\mu_{st} A_\mu(x_s+\frac{a}{2}e_{st}^\mu)}
 e^\nu_{st}\lambda_{\nu}(x_s+\frac{a}{2}e_{st})
 e^{-\frac{i}{2}ae^\mu_{st} A_\mu(x_s+\frac{a}{2}e_{st}^\mu)},\\
 Y_f= a^2 Y(x_f), \quad 
 \chi_f= a^{\frac{3}{2}} \chi(x_f).
\end{array}
 \label{rescale}
\end{equation}
Not only the fields but also the supercharge $Q$ and the coupling constant $g_0$ on the lattice
are dimensionless as well. Therefore they must also be rescaled as 
\begin{equation}
	Q=a^{1/2}\hQ, \qquad \frac{1}{g_0^2}=\frac{1}{a^2g_{2d}^2}. 
 \label{rescale Q}
\end{equation}

Let us now evaluate the action (\ref{lat action}) in the continuum limit. 
Substituting (\ref{rescale}) and (\ref{rescale Q}) in the action (\ref{lat action}),
we obtain
\begin{align}
\label{site action}
S_S&= \frac{\hQ}{2g_{2d}^2} \sum_{f\in F} a^2 A_f
 \left(\sum_{s\in S_f} \frac{\alpha_s^f}{A_f}   
 \Tr \left( \frac{1}{4}\eta(x_s)[\Phi(x_s),\bar\Phi(x_s)]\right)\right), \\
%%%%%%%
\label{link action}
S_L&=\frac{\hQ}{2g_{2d}^2} \sum_{f\in F} a^2 A_f 
\left( \sum_{\link{st}\in L_f} \frac{{\alpha^f}_{\!\!\link{st}}}{A_f} e^\mu_{st} e^\nu_{st} 
\Tr \biggl\{
-i \lambda_\mu(x_s) \cD_\nu \bar\Phi(x_s) + \cO(a) \biggr\}\right),  \\
%%%%%%%
\label{face action}
S_F&= \frac{\hQ}{2g_{2d}^2}\sum_{f\in F} a^2 A_f \left( \frac{\alpha_f}{A_f} 
 \Tr\biggl\{
  \chi(x_f)\left( Y(x_f) - i \beta_f A_f \frac{\epsilon^{\mu\nu}}{\sqrt{g(x_f)}} F_{\mu\nu} + \cO(a)\right)
 \biggr\}\right) , 
\end{align}
where 
$S_f$ is the set of sites that construct the face $f$, 
$F_s$ is the set of faces that meet at the site $s$, 
$\alpha_s^f$ and $\alpha^f_{\link{st}}$ are constants satisfying 
$\alpha_s=\sum_{f\in F_s} \alpha_s^f$ and 
$\alpha_{\link{st}}=\sum_{f\in F_{\link{st}}}  \alpha^f_{\link{st}}$, 
respectively, 
and we have used 
\begin{equation}
\mu(U_f)=ia^2 \frac{A_f}{\sqrt{g(x_f)}} \epsilon^{\mu\nu}F_{\mu\nu}  + \cO(a^3),
\end{equation}
while evaluating 
(\ref{face action}) (see the appendix \ref{proof}). 
Here $F_{\link{st}}$ is the set of faces that share the link $\link{st}$%
\footnote{
If the link $\link{st}$ is a component of the boundary, if it exists, 
of the surface, only one face shares it. 
Otherwise two faces share it. 
}. 
It is easy to see that the continuum limit of 
the site action (\ref{site action}) and the face action (\ref{face action}) 
becomes the corresponding part of the continuum action (\ref{curved action}) 
by setting the parameters $\alpha_s$, $\alpha_f$ and $\beta_f$ as 
\begin{align}
\alpha_s&= \sum_{f\in F_s} \frac{A_f}{|S_f|},\quad
\alpha_f=A_f, \quad \beta_f=\frac{1}{A_f}.
\label{para1}
\end{align}
The link part (\ref{link action}) is slightly more complicated; 
in order to reproduce the continuum action, $\alpha_{\link{st}}$ must satisfy 
\begin{align}
\sum_{\link{st}\in L_f}\alpha^f_\link{st} e^\mu_{st} e^\nu_{st} = A_f g^{\mu\nu}(x_f). 
\label{para2}
\end{align}
It is easy to see that we can determine the value of $\alpha_{\link{st}}$ for any given 
Riemann surface by solving (\ref{para2}). 
In fact, when the face $f$ consists of $n$ links, $l_i$ $(i=1,\cdots,n)$, 
the rank of the $3\times n$ matrix $M^I_i\equiv e^\mu_{l_i} e^\nu_{l_i}$ 
$(I=(\mu,\nu)=(1,1),(2,2),(1,2))$ is three since we assume that all the vectors 
$\vec{e}_{l_i}$ are in the same two-dimensional plane.
In particular, if we consider triangulation, $\alpha^f_{\link{st}}$ are uniquely determined 
through the equation (\ref{para2}). 
Therefore we see that the classical continuum limit of the discretized theory 
(\ref{action}) becomes two-dimensional topological field theory on the Riemann surface $\Sigma_g$ by setting 
$\alpha_s$, $\alpha_{\link{st}}$, $\alpha_f$ and $\beta_f$ as (\ref{para1}) and (\ref{para2}).

% radiative corrections
\subsection{Radiative corrections}
We next discuss possible radiative corrections that appear 
in taking the continuum limit. 
The discussion is completely parallel with that for Sugino's formulation 
given in \cite{Sugino:2003yb,Sugino:2004qd,Sugino:2004uv,Sugino:2006uf}. 
{}From the power counting, we see that possible relevant or marginal operators 
that can appear radiatively are 
$B_1(x)$ or $B_1(x)B_2(x)$ with bosonic fields $B_1(x)$ and $B_2(x)$.
{}From the gauge symmetry and the $\hQ$-symmetry, the only possible terms are
$\Tr\Phi(x)$ and $\Tr\Phi(x)^2$ up to constant factors. 

As announced, 
the situation differs depending on whether the scalar fields $\Phi(x)$ and $\bar\Phi(x)$ are 
complex conjugate with each other or not. 
When $\Phi(x)$ and $\bar\Phi(x)$ are 
complex conjugate with each other as in Sugino's formulation, 
both $\Tr\Phi(x)$ and $\Tr\Phi(x)^2$ are forbidden by the $U(1)_R$ symmetry (\ref{U1-R}). 
Therefore, we do not need any fine-tuning in taking the continuum limit in this case. 
On the other hand, when $\Phi(x)$ and $\bar\Phi(x)$ are independent 
hermitian variables, there is no symmetry that forbids the appearance of these operators radiatively. 
Therefore we need to add counter-terms, 
\begin{equation}
 S_C = 
\begin{cases} 
 \sum_{s\in S} \Tr \left( c_1 \Phi_s^2 + c_2\Phi_s  \right) &{\rm for}\ G=U(N),  \\
 \sum_{s\in S} \Tr\left( c_1 \Phi_s^2 \right) &{\rm for}\ G=SU(N), 
\end{cases}
\end{equation}
to the action and tune the parameters $c_1$ ($c_1$ and $c_2$) for $G=SU(N)$ $(G=U(N))$ 
in taking the continuum limit%
\footnote{Because of the $Q$-symmetry, we see that the expectation values of 
some operators in $Q$-cohomology
can be exactly evaluated even in the lattice theory \cite{OMM}. 
In simulation, therefore, we will be able to use this exact result in tuning $c_1$ and $c_2$. }.

%%%%%%%%%%%%%%%%%%%%%%%%%%%%%%%%%%%%%%%%%%
\section{Conclusion and discussion}

In this paper, we have constructed a discrete formulation of
the topologically twisted $\cN=(2,2)$ supersymmetric Yang-Mills theory 
on an arbitrary two-dimensional lattice while preserving a supercharge. 
When the polygon decomposition (general lattice) is the discretization of the Riemann surface $\Sigma_g$, 
the continuum limit of this theory becomes the topologically twisted $\cN=(2,2)$ 
supersymmetric Yang-Mills theory on $\Sigma_g$. 
If we consider the usual square lattice as an example of the decomposition, 
our model reproduces Sugino's lattice formulation of $\cN=(2,2)$ supersymmetric Yang-Mills theory 
on the torus.

We have also shown that we can take the continuum limit without any fine-tuning if the theory 
possesses $U(1)_R$ symmetry, i.e., we regard the two scalar fields 
in the vector multiplet as being complex conjugate with each other.
On the other hand, if the scalar fields are independent Hermitian variables 
and the gauge group is $SU(N)$ (or $U(N)$), 
there is no $U(1)_R$ symmetry in the model and we need one-parameter 
(or two-parameters) tuning in the continuum limit.

A natural question would arise as to whether there is a fermion doubler in this model or not. 
In order to answer this question, we have to examine if the kinetic terms of the fermions 
have no non-trivial zero, which depends on the structure of the discretization. 
However, we should recall that the origin of the fermion doubler is the periodicity 
in the momentum space, 
which is associated with the discrete translational invariance of lattice. 
Since a general lattice has less discrete translational symmetry 
than the usual square lattice, there is less chance for fermion doublers to appear. 
In addition, even if we consider the square lattice, 
it is shown that fermion doubler is absent \cite{Sugino:2003yb}.
Although it is still possible that fermion doublers appear by discretizing 
the Riemann surface by a highly symmetric tiling, 
we can conclude that there is no fermion doubler in most cases.

In the continuum theory, the so-called localization is used to examine the topological 
nature of the two-dimensional gauge theory \cite{Witten:1992xu}. 
Since our model preserves the scalar supersymmetry, which is the crucial symmetry 
in order that localization works, we can use the same technique in the discretized theory,
which will be discussed separately in \cite{OMM}. 

It will be straightforward to apply our method to the two-dimensional $\cN=(4,4)$ and 
$(8,8)$ supersymmetric Yang-Mills theories or two-dimensional supersymmetric QCD. 
{}Furthermore our method is also applicable to 
the orbifold lattice theory \cite{Kaplan:2002wv,Cohen:2003xe,Cohen:2003qw,Kaplan:2005ta}. 
The original orbifold lattice theory is based on the concept of deconstruction and  
is constructed by dividing a matrix theory (mother theory) 
by a discrete subgroup of the mother theory. 
The only background we can obtain in this way is the torus: it seems to be impossible 
that the standard orbifold projection constructs
a theory on an arbitrary Riemann surface.
On the other hand, by using our method, we can construct the theory on the arbitrary lattice 
and we can embed the fields in sparse matrices. 
In this sense, our method can be regarded as a non-trivial extension of deconstruction, 
which will be connected with network theory. 
It might provide a novel way to examine the topological nature of gauge theory. 

Including the fluctuation of polygons like 
Regge calculus \cite{Regge:1961px}
or
dynamical triangulation \cite{2DDT} 
will be a fascinating next step. 
To this end, our set-up given in the section 3 would be insufficient to 
generate Riemann surface dynamically because it includes too-wide discretized objects. 
One plausible idea is to restrict the discretization to a simplicial complex. 
It will be interesting question to see if the diffeomorphism invariance is recovered 
in the continuum limit under such a restriction.

\section*{Acknowledgements}
The authors would like to thank
N.~Kawamoto,
K.~Murata,
N.~Sakai,
F.~Sugino,
H.~Suzuki, 
T.~Tada,
and Y.~Yoshida 
for useful discussions. 
The work of S.M., T.M. and K.O. was supported in part by a Grant-in-Aid
for Young Scientists (B), 23740197,
a Grant-in-Aid
for Young Scientists (B), 26800417,
and JSPS KAKENHI Grant Number 14485514, respectively.

\appendix
\section{Continuum limit of the plaquette variable}
\label{proof}

Let us consider a face $(s_1,\cdots,s_n)$ and the corresponding plaquette variable, 
\begin{equation}
 U_f=\prod_{i=1}^n U_{s_is_{i+1}} \quad (s_{n+1}=s_1).
 \label{A1:Uf}
\end{equation}
We here assume that the vectors $e_{s_ks_{k+1}}$ constructing this face 
span the same two-dimensional plane. 
Recalling that it is reasonable to think that the continuum gauge field 
is living at the middle point of the link: 
\begin{equation}
 U_{s_k s_{k+1}}=\exp\left\{ i a e^\mu_{s_k s_{k+1}} A_\mu ( s_k + \frac{a}{2} e_{s_k s_{k+1}} ) \right\}, 
 \label{A1:U to A}
\end{equation}
and the argument of $A_\mu$ is rewritten as 
\begin{equation}
 s_k + \frac{a}{2} e_{s_k s_{k+1}} 
 = s_1 + \frac{a}{2} \left(
 e_{s_1s_2}+e_{s_2s_3}+\cdots+e_{s_{k-1}s_k}-e_{s_{k+1}s_{k+2}}-\cdots -e_{s_ns_1}
 \right), 
\end{equation}
we can rewrite (\ref{A1:U to A}) as 
\begin{equation}
U_{s_ks_{k+1}}=\exp\left\{ ia e^\mu_{st} A_\mu(s_1)
+ \frac{i}{2}a^2 e^\mu_{s_k s_{k+1}} 
\left( \sum_{l<k} e^\nu_{s_l s_{l+1}} - \sum_{l>k} e^\nu_{s_l s_{l+1}} \right)
\del_\nu A_\mu(s_1)
+ \cO(a^3) \right\}. 
\label{A1:expand}
\end{equation}
Substituting (\ref{A1:expand}) to (\ref{A1:Uf}) and 
using the Campbell-Baker-Hausdorff formula, 
\begin{equation}
e^{M_1}e^{M_2}\cdots e^{M_n} = e^{\sum_{i=1}^n M_i + \frac{1}{2} \sum_{i<j}[M_i,M_j] + \cdots }, 
\end{equation}
we see 
\begin{equation}
U_f = \exp\left\{ \frac{i}{2} a^2 C_f^{\mu\nu} F_{\mu\nu}(s_1) + \cO(a^3) \right\}, 
\end{equation}
where 
\begin{equation}
F_{\mu\nu}=\del_\mu A_\nu - \del_\nu A_\mu + i[A_\mu, A_\nu] , 
\end{equation}
and 
\begin{equation}
 C^{\mu\nu}_f = \frac{1}{2} \sum_{k=1}^n e^\mu_{s_k s_{k+1}}
 \left(
- \sum_{l<k} e^\nu_{s_l s_{l+1}} + \sum_{l>k} e^\nu_{s_l s_{l+1}}
 \right). 
\end{equation}

In order to see the geometrical meaning of $C^{\mu\nu}_f$, 
it is convenient to rewrite it as 
\begin{equation}
C^{\mu\nu}_f = \frac{1}{2} \sum_{i=3}^n \left(
e^\mu_{s_i s_{i-1}} e^\nu_{s_i s_1} - e^\nu_{s_i s_{i-1}} e^\mu_{s_i s_1} 
\right), 
\end{equation}
where $e_{s_is_1}\equiv -e_{s_i s_i+1} - e_{s_{i+1} s_{i+2}} - \cdots e_{s_n s_1}$. 
Since 
$\frac{1}{2}  (e^1_{s_i s_{i-1}} e^2_{s_i s_1} - e^2_{s_i s_{i-1}} e^1_{s_i s_1})$ 
is the area of the triangle with the vertices $s_1, s_{i-1}, s_{i}$, 
we see 
\begin{align}
%C_f^{\mu\nu} &= [\text{area of the polygon made up of } e_{s_i s_{i+1}}]\, \epsilon^{\mu\nu} 
C_f^{\mu\nu}=\frac{A_f}{\sqrt{g(x_f)}} \epsilon^{\mu\nu},
\end{align}
which is proportional to a unit area of the polygon made up of $e_{s_i s_{i+1}}$'s.

\providecommand{\href}[2]{#2}\begingroup\raggedright\endgroup

% partial difference
%\section{More on the link action}
%\label{partial}
%\bibliographystyle{JHEP}
%\bibliography{refs.bib}

\begin{thebibliography}{10}

\bibitem{Elitzur:1982vh}
S.~Elitzur, E.~Rabinovici and A.~Schwimmer, {\it SUPERSYMMETRIC MODELS ON THE
  LATTICE},  {\em Phys. Lett.} {\bf B119} (1982) 165.
%%CITATION = PHLTA,B119,165;%%

\bibitem{Banks:1982ut}
T.~Banks and P.~Windey, {\it {SUPERSYMMETRIC LATTICE THEORIES}},  {\em
  Nucl.Phys.} {\bf B198} (1982) 226--236.
%%CITATION = NUPHA,B198,226;%%

\bibitem{Ichinose:1982ug}
I.~Ichinose, {\it {SUPERSYMMETRIC LATTICE GAUGE THEORY}},  {\em Phys.Lett.}
  {\bf B122} (1983) 68.
%%CITATION = PHLTA,B122,68;%%

\bibitem{Bartels:1983wm}
J.~Bartels and J.~Bronzan, {\it {SUPERSYMMETRY ON A LATTICE}},  {\em Phys.Rev.}
  {\bf D28} (1983) 818.
%%CITATION = PHRVA,D28,818;%%

\bibitem{Kaplan:2002wv}
D.~B. Kaplan, E.~Katz and M.~Unsal, {\it Supersymmetry on a spatial lattice},
  {\em JHEP} {\bf 05} (2003) 037
  [\href{http://arXiv.org/abs/hep-lat/0206019}{{\tt hep-lat/0206019}}].
%%CITATION = HEP-LAT 0206019;%%

\bibitem{Cohen:2003xe}
A.~G. Cohen, D.~B. Kaplan, E.~Katz and M.~Unsal, {\it Supersymmetry on a
  Euclidean spacetime lattice. I: A target theory with four supercharges},
  {\em JHEP} {\bf 08} (2003) 024
  [\href{http://arXiv.org/abs/hep-lat/0302017}{{\tt hep-lat/0302017}}].
%%CITATION = HEP-LAT 0302017;%%

\bibitem{Cohen:2003qw}
A.~G. Cohen, D.~B. Kaplan, E.~Katz and M.~Unsal, {\it Supersymmetry on a
  Euclidean spacetime lattice. II: Target theories with eight supercharges},
  {\em JHEP} {\bf 12} (2003) 031
  [\href{http://arXiv.org/abs/hep-lat/0307012}{{\tt hep-lat/0307012}}].
%%CITATION = HEP-LAT 0307012;%%

\bibitem{Kaplan:2005ta}
D.~B. Kaplan and M.~Unsal, {\it A Euclidean lattice construction of
  supersymmetric Yang- Mills theories with sixteen supercharges},  {\em JHEP}
  {\bf 09} (2005) 042 [\href{http://arXiv.org/abs/hep-lat/0503039}{{\tt
  hep-lat/0503039}}].
%%CITATION = HEP-LAT 0503039;%%

\bibitem{Endres:2006ic}
M.~G. Endres and D.~B. Kaplan, {\it Lattice formulation of (2,2) supersymmetric
  gauge theories with matter fields},  {\em JHEP} {\bf 10} (2006) 076
  [\href{http://arXiv.org/abs/hep-lat/0604012}{{\tt hep-lat/0604012}}].
%%CITATION = HEP-LAT 0604012;%%

\bibitem{Giedt:2006dd}
J.~Giedt, {\it {Quiver lattice supersymmetric matter, D1/D5 branes and
  AdS(3)/CFT(2)}},  \href{http://arXiv.org/abs/hep-lat/0605004}{{\tt
  hep-lat/0605004}}.
%%CITATION = HEP-LAT/0605004;%%

\bibitem{Matsuura:2008cfa}
S.~Matsuura, {\it {Two-dimensional N=(2,2) Supersymmetric Lattice Gauge Theory
  with Matter Fields in the Fundamental Representation}},  {\em JHEP} {\bf
  0807} (2008) 127 [\href{http://arXiv.org/abs/0805.4491}{{\tt 0805.4491}}].
%%CITATION = ARXIV:0805.4491;%%

\bibitem{Catterall:2003wd}
S.~Catterall, {\it Lattice supersymmetry and topological field theory},  {\em
  JHEP} {\bf 05} (2003) 038 [\href{http://arXiv.org/abs/hep-lat/0301028}{{\tt
  hep-lat/0301028}}].
%%CITATION = HEP-LAT 0301028;%%

\bibitem{DAdda:2004jb}
A.~D'Adda, I.~Kanamori, N.~Kawamoto and K.~Nagata, {\it Twisted superspace on a
  lattice},  {\em Nucl. Phys.} {\bf B707} (2005) 100--144
  [\href{http://arXiv.org/abs/hep-lat/0406029}{{\tt hep-lat/0406029}}].
%%CITATION = HEP-LAT 0406029;%%

\bibitem{Nagata:2008zz}
K.~Nagata and Y.-S. Wu, {\it {Twisted SUSY Invariant Formulation of
  Chern-Simons Gauge Theory on a Lattice}},
  \href{http://arXiv.org/abs/0803.4339}{{\tt 0803.4339}}.
%%CITATION = 0803.4339;%%

\bibitem{Joseph:2013jya}
A.~Joseph, {\it {Lattice formulation of three-dimensional ${\cal N}=4$ gauge
  theory with fundamental matter fields}},  {\em JHEP} {\bf 1309} (2013) 046
  [\href{http://arXiv.org/abs/1307.3281}{{\tt 1307.3281}}].
%%CITATION = ARXIV:1307.3281;%%

\bibitem{Unsal:2006qp}
M.~Unsal, {\it Twisted supersymmetric gauge theories and orbifold lattices},
  {\em JHEP} {\bf 10} (2006) 089
  [\href{http://arXiv.org/abs/hep-th/0603046}{{\tt hep-th/0603046}}].
%%CITATION = HEP-TH 0603046;%%

\bibitem{Catterall:2007kn}
S.~Catterall, {\it {From Twisted Supersymmetry to Orbifold Lattices}},  {\em
  JHEP} {\bf 01} (2008) 048 [\href{http://arXiv.org/abs/0712.2532}{{\tt
  0712.2532}}].
%%CITATION = 0712.2532;%%

\bibitem{Damgaard:2007eh}
P.~H. Damgaard and S.~Matsuura, {\it Lattice Supersymmetry: Equivalence between
  the Link Approach and Orbifolding},  {\em JHEP} {\bf 09} (2007) 097
  [\href{http://arXiv.org/abs/0708.4129 [hep-lat]}{{\tt 0708.4129
  [hep-lat]}}].
%%CITATION = ARXIV:0708.4129;%%

\bibitem{Catterall:2009it}
S.~Catterall, D.~B. Kaplan and M.~Unsal, {\it {Exact lattice supersymmetry}},
  {\em Phys.Rept.} {\bf 484} (2009) 71--130
  [\href{http://arXiv.org/abs/0903.4881}{{\tt 0903.4881}}].
%%CITATION = ARXIV:0903.4881;%%

\bibitem{Kanamori:2008bk}
I.~Kanamori and H.~Suzuki, {\it {Restoration of supersymmetry on the lattice:
  Two-dimensional N = (2,2) supersymmetric Yang-Mills theory}},  {\em
  Nucl.Phys.} {\bf B811} (2009) 420--437
  [\href{http://arXiv.org/abs/0809.2856}{{\tt 0809.2856}}].
%%CITATION = ARXIV:0809.2856;%%

\bibitem{Hanada:2009hq}
M.~Hanada and I.~Kanamori, {\it {Lattice study of two-dimensional N=(2,2) super
  Yang-Mills at large-N}},  {\em Phys.Rev.} {\bf D80} (2009) 065014
  [\href{http://arXiv.org/abs/0907.4966}{{\tt 0907.4966}}].
%%CITATION = ARXIV:0907.4966;%%

\bibitem{Catterall:2010fx}
S.~Catterall, A.~Joseph and T.~Wiseman, {\it {Thermal phases of D1-branes on a
  circle from lattice super Yang-Mills}},  {\em JHEP} {\bf 1012} (2010) 022
  [\href{http://arXiv.org/abs/1008.4964}{{\tt 1008.4964}}].
%%CITATION = ARXIV:1008.4964;%%

\bibitem{Sugino:2003yb}
F.~Sugino, {\it A lattice formulation of super Yang-Mills theories with exact
  supersymmetry},  {\em JHEP} {\bf 01} (2004) 015
  [\href{http://arXiv.org/abs/hep-lat/0311021}{{\tt hep-lat/0311021}}].
%%CITATION = HEP-LAT 0311021;%%

\bibitem{Sugino:2004qd}
F.~Sugino, {\it Super Yang-Mills theories on the two-dimensional lattice with
  exact supersymmetry},  {\em JHEP} {\bf 03} (2004) 067
  [\href{http://arXiv.org/abs/hep-lat/0401017}{{\tt hep-lat/0401017}}].
%%CITATION = HEP-LAT 0401017;%%

\bibitem{Sugino:2004uv}
F.~Sugino, {\it Various super Yang-Mills theories with exact supersymmetry on
  the lattice},  {\em JHEP} {\bf 01} (2005) 016
  [\href{http://arXiv.org/abs/hep-lat/0410035}{{\tt hep-lat/0410035}}].
%%CITATION = HEP-LAT 0410035;%%

\bibitem{Sugino:2006uf}
F.~Sugino, {\it Two-dimensional compact N = (2,2) lattice super Yang-Mills
  theory with exact supersymmetry},  {\em Phys. Lett.} {\bf B635} (2006)
  218--224 [\href{http://arXiv.org/abs/hep-lat/0601024}{{\tt
  hep-lat/0601024}}].
%%CITATION = HEP-LAT 0601024;%%

\bibitem{Sugino:2008yp}
F.~Sugino, {\it {Lattice Formulation of Two-Dimensional N=(2,2) SQCD with Exact
  Supersymmetry}},  {\em Nucl.Phys.} {\bf B808} (2009) 292--325
  [\href{http://arXiv.org/abs/0807.2683}{{\tt 0807.2683}}].
%%CITATION = ARXIV:0807.2683;%%

\bibitem{Kikukawa:2008xw}
Y.~Kikukawa and F.~Sugino, {\it {Ginsparg-Wilson Formulation of 2D N = (2,2)
  SQCD with Exact Lattice Supersymmetry}},  {\em Nucl.Phys.} {\bf B819} (2009)
  76--115 [\href{http://arXiv.org/abs/0811.0916}{{\tt 0811.0916}}].
%%CITATION = ARXIV:0811.0916;%%

\bibitem{Suzuki:2007jt}
H.~Suzuki, {\it {Two-dimensional $\mathcal{N}=(2,2)$ super Yang-Mills theory on
  computer}},  {\em JHEP} {\bf 09} (2007) 052
  [\href{http://arXiv.org/abs/0706.1392}{{\tt 0706.1392}}].
%%CITATION = 0706.1392;%%

\bibitem{Kanamori:2007ye}
I.~Kanamori, H.~Suzuki and F.~Sugino, {\it {Euclidean lattice simulation for
  the dynamical supersymmetry breaking}},
  \href{http://arXiv.org/abs/0711.2099}{{\tt 0711.2099}}.
%%CITATION = 0711.2099;%%

\bibitem{Kanamori:2007yx}
I.~Kanamori, F.~Sugino and H.~Suzuki, {\it {Observing dynamical supersymmetry
  breaking with euclidean lattice simulations}},
  \href{http://arXiv.org/abs/0711.2132}{{\tt 0711.2132}}.
%%CITATION = 0711.2132;%%

\bibitem{Matsuura:2014pua}
S.~Matsuura and F.~Sugino, {\it {Lattice formulation for 2d = (2, 2), (4, 4)
  super Yang-Mills theories without admissibility conditions}},  {\em JHEP}
  {\bf 1404} (2014) 088 [\href{http://arXiv.org/abs/1402.0952}{{\tt
  1402.0952}}].
%%CITATION = ARXIV:1402.0952;%%

\bibitem{Maru:1997kh}
N.~Maru and J.~Nishimura, {\it Lattice formulation of supersymmetric Yang-Mills
  theories without fine-tuning},  {\em Int. J. Mod. Phys.} {\bf A13} (1998)
  2841--2856 [\href{http://arXiv.org/abs/hep-th/9705152}{{\tt
  hep-th/9705152}}].
%%CITATION = HEP-TH/9705152;%%

\bibitem{Giedt:2009yd}
J.~Giedt, {\it {Progress in four-dimensional lattice supersymmetry}},  {\em
  Int.J.Mod.Phys.} {\bf A24} (2009) 4045--4095
  [\href{http://arXiv.org/abs/0903.2443}{{\tt 0903.2443}}].
%%CITATION = ARXIV:0903.2443;%%



\bibitem{Berenstein:2002jq}
D.~E. Berenstein, J.~M. Maldacena and H.~S. Nastase, {\it {Strings in flat
  space and pp waves from N=4 superYang-Mills}},  {\em JHEP} {\bf 0204} (2002)
  013 [\href{http://arXiv.org/abs/hep-th/0202021}{{\tt hep-th/0202021}}].
%%CITATION = HEP-TH/0202021;%%

\bibitem{Das:2003yq}
S.~R. Das, J.~Michelson and A.~D. Shapere, {\it {Fuzzy spheres in pp wave
  matrix string theory}},  {\em Phys.Rev.} {\bf D70} (2004) 026004
  [\href{http://arXiv.org/abs/hep-th/0306270}{{\tt hep-th/0306270}}].
%%CITATION = HEP-TH/0306270;%%

\bibitem{Myers:1999ps}
R.~C. Myers, {\it {Dielectric branes}},  {\em JHEP} {\bf 9912} (1999) 022
  [\href{http://arXiv.org/abs/hep-th/9910053}{{\tt hep-th/9910053}}].
%%CITATION = HEP-TH/9910053;%%

\bibitem{Ishii:2008ib}
T.~Ishii, G.~Ishiki, S.~Shimasaki and A.~Tsuchiya, {\it {N=4 Super Yang-Mills
  from the Plane Wave Matrix Model}},  {\em Phys.Rev.} {\bf D78} (2008) 106001
  [\href{http://arXiv.org/abs/0807.2352}{{\tt 0807.2352}}].
%%CITATION = ARXIV:0807.2352;%%

\bibitem{Ishiki:2008te}
G.~Ishiki, S.~W.~Kim, J.~Nishimura and A.~Tsuchiya,
{\it Deconfinement phase transition in N=4 super Yang-Mills theory on $R\times S^3$ from
supersymmetric matrix quantum mechanics}, 
{\em Phys.Rev.Lett.} {\bf 102} (2009) 111601
  [\href{http://arXiv.org/abs/0810.2884}{{\tt 0810.2884}}].
%%CITATION = PRLTA,102,111601;%%

\bibitem{Ishiki:2009sg}
G.~Ishiki, S.~W.~Kim, J.~Nishimura and A.~Tsuchiya,
{\it Testing a novel large-N reduction for N=4 super Yang-Mills theory on $R\times S^3$},
{\em JHEP} {\bf 0909} (2009) 029
  [\href{http://arXiv.org/abs/0907.1488}{{\tt 0907.1488}}]
%%CITATION = JHEPA,0909,029;%%

\bibitem{Ishiki:2011ct}
G.~Ishiki, S.~Shimasaki and A.~Tsuchiya, {\it {Perturbative tests for a large-N
  reduced model of super Yang-Mills theory}},  {\em JHEP} {\bf 1111} (2011) 036
  [\href{http://arXiv.org/abs/1106.5590}{{\tt 1106.5590}}].
%%CITATION = ARXIV:1106.5590;%%

\bibitem{Honda:2013nfa}
M.~Honda, G.~Ishiki, S.~W.~Kim, J.~Nishimura and A.~Tsuchiya,
{\it Direct test of the AdS/CFT correspondence by Monte Carlo studies 
of N=4 super Yang-Mills theory},
{\em JHEP} {\bf 1311} (2013) 200
  [\href{http://arXiv.org/abs/1308.3525}{{\tt 1308.3525}}]
%%CITATION = ARXIV:1308.3525;%%

\bibitem{Hanada:2010kt}
M.~Hanada, S.~Matsuura and F.~Sugino, {\it {Two-dimensional lattice for
  four-dimensional N=4 supersymmetric Yang-Mills}},  {\em Prog.Theor.Phys.}
  {\bf 126} (2011) 597--611 [\href{http://arXiv.org/abs/1004.5513}{{\tt
  1004.5513}}].
%%CITATION = ARXIV:1004.5513;%%

\bibitem{Hanada:2010gs}
M.~Hanada, {\it {A proposal of a fine tuning free formulation of 4d N = 4 super
  Yang-Mills}},  {\em JHEP} {\bf 1011} (2010) 112
  [\href{http://arXiv.org/abs/1009.0901}{{\tt 1009.0901}}].
%%CITATION = ARXIV:1009.0901;%%

\bibitem{Hanada:2011qx}
M.~Hanada, S.~Matsuura and F.~Sugino, {\it {Non-perturbative construction of 2D
  and 4D supersymmetric Yang-Mills theories with 8 supercharges}},  {\em
  Nucl.Phys.} {\bf B857} (2012) 335--361
  [\href{http://arXiv.org/abs/1109.6807}{{\tt 1109.6807}}].
%%CITATION = ARXIV:1109.6807;%%

\bibitem{Catterall:N=41}
S. Catterall, E. Dzienkowski, J. Giedt, A. Joseph, and R. Wells, 
{\it Perturbative renormalization of lattice N=4 super Yang-Mills theory}, 
{\em JHEP} {\bf 1104} (2011) 074, [{\tt 1102.1725}].

\bibitem{Catterall:N=42}
S. Catterall, P. H. Damgaard, T. Degrand, R. Galvez and D. Mehta, 
 {\it Phase Structure of Lattice N=4 Super Yang-Mills}, 
 {\em JHEP} {\bf 1211} (2012) 072 [{\tt 1209.5285}]. 
 
 
 \bibitem{Catterall:N=43}
S. Catterall, J. Giedt, and A. Joseph, 
 {\it Twisted supersymmetries in lattice N = 4 super Yang-Mills theory}, 
  {\em JHEP} {\bf 1310} (2013) 166, [{\tt 1306.3891}].
  
\bibitem{Catterall:N=44}
S.~Catterall, D.~Schaich, P.~H.~Damgaard, T.~DeGrand and J.~Giedt,
  {\it N=4 Supersymmetry on a Space-Time Lattice},
  {\em Phys.Rev.} {\bf D90} (2014) 065013   [{\tt 1405.0644}], 
  %%CITATION = ARXIV:1405.0644;%%
  
\bibitem{Catterall:N=45}
S.~Catterall and J.~Giedt,
 {\it Real space renormalization group for twisted lattice N=4 super Yang-Mills},
  [{\tt 1408.7067}]. 
  %%CITATION = ARXIV:1408.7067;%%

\bibitem{Witten:1988ze}
E.~Witten, {\it Topological Quantum Field Theory},  {\em Commun. Math. Phys.}
  {\bf 117} (1988) 353.
%%CITATION = CMPHA,117,353;%%

\bibitem{Pestun:2007rz}
V.~Pestun, {\it {Localization of gauge theory on a four-sphere and
  supersymmetric Wilson loops}},  {\em Commun.Math.Phys.} {\bf 313} (2012)
  71--129 [\href{http://arXiv.org/abs/0712.2824}{{\tt 0712.2824}}].
%%CITATION = ARXIV:0712.2824;%%

\bibitem{Witten:1990bs}
E.~Witten, {\it Introduction to cohomological field theories},  {\em Int. J.
  Mod. Phys.} {\bf A6} (1991) 2775--2792.
%%CITATION = IMPAE,A6,2775;%%

\bibitem{Bershadsky:1995qy}
M.~Bershadsky, C.~Vafa and V.~Sadov, {\it {D-branes and topological field
  theories}},  {\em Nucl.Phys.} {\bf B463} (1996) 420--434
  [\href{http://arXiv.org/abs/hep-th/9511222}{{\tt hep-th/9511222}}].
%%CITATION = HEP-TH/9511222;%%

\bibitem{Witten:1992xu}
E.~Witten, {\it {Two-dimensional gauge theories revisited}},  {\em
  J.Geom.Phys.} {\bf 9} (1992) 303--368
  [\href{http://arXiv.org/abs/hep-th/9204083}{{\tt hep-th/9204083}}].
%%CITATION = HEP-TH/9204083;%%

\bibitem{OMM}
K.~Ohta, S.~Matsuura and T.~Misumi, {\it in preparation}.

%\cite{Ambjorn:1991pq}
%\bibitem{Ambjorn:1991pq} 
%  J.~Ambjorn and J.~Jurkiewicz,
%  {\it Four-dimensional simplicial quantum gravity}, 
%  {\em Phys. Lett. B} {\bf 278} (1992) 42. 
  %%CITATION = PHLTA,B278,42;%%
  %154 citations counted in INSPIRE as of 19 Sep 2014

%\cite{Agishtein:1991cv}
%\bibitem{Agishtein:1991cv} 
%  M.~E.~Agishtein and A.~A.~Migdal,
%  {\it Simulations of four-dimensional simplicial quantum gravity}, 
%  {\em Mod. Phys. Lett.}  {\bf A7} (1992) 1039, 
%  {\em Nucl. Phys.} {\bf B385} (1992) 395.
  %%CITATION = MPLAE,A7,1039;%%
  %111 citations counted in INSPIRE as of 19 Sep 2014

%\cite{Regge:1961px}
\bibitem{Regge:1961px} 
  T.~Regge,
  {\it General Relativity Without Coordinates}, 
  {\em Nuovo Cim.}  {\bf 19} (1961) 558 .
  %%CITATION = NUCIA,19,558;%%
  %606 citations counted in INSPIRE as of 19 Sep 2014

\bibitem{2DDT}
D.~Weingarten, 
{\it Euclidean Quantum Gravity on a Lattice}, 
{\em Nucl. Phys. } {\bf B210} (1982), 229, 
J.~Ambjorn, B.~Durhuus and J.~Frohlich,
{\it Diseases of Triangulated Random Surface Models, and Possible Cures,} 
{\em Nucl. Phys.}  {\bf B257} (1985) 433, 
  %%CITATION = NUPHA,B257,433;%%
F.~David,
{\it Planar Diagrams, Two-Dimensional Lattice Gravity and Surface Models,}
{\em Nucl. Phys.} {\bf B257} (1985) 45, 
  %%CITATION = NUPHA,B257,45;%%
V.~A.~Kazakov, A.~A.~Migdal and I.~K.~Kostov,
{\it Critical Properties of Randomly Triangulated Planar Random Surfaces,}
{\em Phys. Lett.}  {\bf B157} (1985) 295.
  %%CITATION = PHLTA,B157,295;%%

\end{thebibliography}
%\providecommand{\href}[2]{#2}\begingroup\raggedright\begin{thebibliography}    {10}
%\end{thebibliography}\endgroup

\end{document}